\RequirePackage{fix-cm}
\documentclass[smallextended]{svjour3}       
\usepackage[italian,english]{babel}
\smartqed  
\usepackage{amsmath}
\usepackage{amssymb}
\usepackage{lipsum}
\usepackage{dsfont}
\usepackage{multirow}
\usepackage{afterpage}
\usepackage{amsfonts}
\usepackage{bm}%
\usepackage{babel}
\usepackage{hyperref}
\usepackage{graphicx}
\usepackage{color}
\usepackage{graphicx}
\begin{document}

\title{Intrinsic decoherence effects on measurement--induced nonlocality 
}


\author{R. Muthuganesan        \and
       V. K. Chandrasekar 
}


\institute{Center for Nonlinear Science \& Engineering, School of Electrical \& Electronics Engineering,
SASTRA Deemed University, Thanjavur, Tamil Nadu 613 401, India\at
              Tel.: +91-8760372825\\
              \email{rajendramuthu@gmail.com}           
           \and
         Center for Nonlinear Science \& Engineering, School of Electrical \& Electronics Engineering,
SASTRA Deemed University, Thanjavur, Tamil Nadu 613 401, India\at
              \email{chandru@gmail.com}
}

\date{Received: date / Accepted: date}

\maketitle

\begin{abstract}
By considering an exactly solvable model of a  two interacting spin-1/2 qubits described by the Heisenberg anisotropic interaction in the presence of intrinsic decoherence, we study the dynamics of entanglement quantified by the concurrence and measurement-induced nonlocality (MIN) based on Hilbert-Schmidt norm and trace distance with different initial conditions. We highlight the relationship between the entanglement and MIN for the pure initial state. For an initial separable state, it is found that the robustness and the generation of the quantum correlations depend on the physical parameters. While considering the entangled state as an initial state, the results show that despite the phase decoherence all the correlations reach their steady state values after exhibiting some oscillations. We reveal that the enhancement of correlations may occur by adjusting the strength of the Dzyaloshinskii--Moriya (DM) interaction and the intervention of the magnetic field decrease the quantum correlations. Finally, we show that the existence of quantum correlation captured by MIN in the unentangled state.

\end{abstract}

\section{Introduction}
\label{intro}
For many decades, entanglement \cite{Einstein,Schrodinger} is considered as a synonym of quantum correlation and it is a direct consequence of the superposition principle. Quantum entanglement,  a special type of correlation that only arises in quantum systems,  along with the superposition principle, stimulates the advances in the field of quantum technology. It reflects nonlocal distributions between pairs of particles, even if they are spatially separated and do not directly interact with each other. In the early days of quantum information theory, entanglement is viewed as the main resource that gives the speed-up over their classical counterparts \cite{Nielsen}. Later, it is identified that even separable state is also responsible for quantum advantages. For example, Knill and Laflamme discovered a protocol of deterministic quantum computation with one quantum bit (DQC1) where the natural bipartite system is unentangled \cite{Knill}. It can achieve an exponential efficiency over classical computers, for a limited set of tasks \cite{Datta2005,Datta2005PRL}. The Knill-Laflamme model is experimentally motivated by (liquid-state) nuclear-magnetic resonance (NMR) information processing. This started to throw doubt on entanglement being responsible for all quantum speed-up and leads to the quest for quantum correlations beyond entanglement.

In 2001 while analysing different measures of information content in quantum information theory, Henderson and Vedral \cite{Henderson}, and (independently) Ollivier and Zurek \cite{Ollivier} conclude that when entanglement is subtracted from total quantum correlation, there remain correlations that are not entirely classical of origin. This is named as quantum discord (QD). Then scientists tried to connect the discord with the performance of certain information processing tasks. Datta \cite{Datta2005} calculated the discord in the Knill-Laflamme algorithm and show the increase in quantum efficiency, unlike entanglement which remains vanishingly small throughout the computation.

There are various correlation measures are available to capture beyond entanglement, such as geometric  Discord \cite{Gediscord}, measurement–induced nonlocality (MIN)\cite{MIN}, measurement–induced disturbance (MID) \cite{MID}, uncertainty-induced nonlocality (UIN) \cite{UIN}, etc. MIN is a measure of bipartite quantum correlation  and  maximal nonlocal effects due to local projective measurements. This quantity is a more secure resource for quantum communication and cryptography than the entanglement. This quantity is easily computable and also experimentally realizable. However, MIN is not a bonafide measure to capture quantum correlation due to the local ancilla problem \cite{Piani2012,Chang2013}. To address this issue different forms of MINs have been identified using relative entropy \cite{RMIN}, trace distance \cite{HUTMIN}, fidelity \cite{FMIN}, skew information \cite{skewMIN}, and affinity \cite{Affinity}.  

In general, quantum systems are always coupled with the environment. The interaction of systems with their environments,  another  dynamics  of  an  open  quantum  system  is  characterized by the presence of decoherence terms, describing the  loss  of  energy,  coherence,  and  information  into  the environments. In order to perform efficient information processing, we have to protect the quantum resources such as entanglement against external perturbation or decoherence. Due to the decoherence parameter, the entanglement completely vanishes and is known as the sudden death of entanglement. Independently, Milburn \cite{Milburn} and Moya-Cessa et. al. \cite{Moya} demonstrated that the quantum signatures are automatically destroyed as the quantum system evolves.  This process is considered to be more generic to nature and which causes phase decoherence even without any averaging over an environment. This is known as the intrinsic decoherence approach. Therefore the study on intrinsic decoherence has profound importance. 

In this present submission, we study the influence of intrinsic decoherence on the dynamics of pair of spin-1/2 qubit quantum correlations such as entanglement (measured by concurrence) and MIN (based on Hilbert-Schmidt norm and trace distance). It is observed that the pure state quantum correlation measures are wiggles and in the asymptotic limit, the state evolves into a steady state. The observation highlights the MINs are more robust than the entanglement against intrinsic decoherence. 

The contents of this paper are as follows.  In Sec. \ref{correlation}, we define the quantifier of quantum correlation studied in this paper. In Sec. \ref{model}, we introduce the theoretical model under our investigation and the notion of intrinsic decoherence in a quantum system. The dynamics of entanglement and MINs presented in Sec. \ref{Dynamics}. Finally, the conclusions are given in Sec. \ref{Concl}. 

\section{Quantum correlation measures}

\label{correlation}

\noindent \textit{Entanglement}: Let  $\rho$ be a bipartite composite system shared by the subsystem $a$ and $b$ in the Hilbert space $\mathcal{H}=\mathcal{H}^a\otimes\mathcal{H}^b$. The degree of entanglement associated with a given two-qubit state $\rho$ can be quantified using concurrence \cite{Wootters}, which is defined as 
\begin{equation}
C(\rho )=\text{max}\{0,~\lambda_1-\lambda_2-\lambda_3-\lambda_4\},
\end{equation}
where $\lambda_i$ are the square root of eigenvalues of matrix $R=\rho \tilde{\rho }$ arranged in decreasing order. Here $\tilde{\rho } $ is spin flipped density matrix, which is defined as $\tilde{\rho }=(\sigma _y \otimes \sigma _y)\rho^{*}(\sigma _y \otimes \sigma _y)$. The symbol $*$ denotes the usual complex conjugate in the computational basis. It is known that the concurrence varies from $0$ to $1$ with minimum and maximum values correspond to separable and maximally entangled states respectively.

\noindent \textit{Measurement induced nonlocality}:
This measure of quantum correlations (QC) is defined as the maximal distance between the quantum state of a bipartite system and the corresponding state after performing a local measurement on one subsystem which does not change the state of this subsystem. This quantity also captures the nonlocal effects that can be induced by local measurements. Mathematically it is defined as \cite{MIN}
\begin{equation}
 N_2(\rho ) =~^{\text{max}}_{\Pi ^{a}}\| \rho - \Pi ^{a}(\rho )\| ^{2},
\end{equation}
where the maximum is taken over the von Neumann projective measurements on subsystem $a$. Here $\Pi^{a}(\rho) = \sum _{k} (\Pi ^{a}_{k} \otimes   \mathds{1} ^{b}) \rho (\Pi ^{a}_{k} \otimes    \mathds{1}^{b} )$, with $\Pi ^{a}= \{\Pi ^{a}_{k}\}= \{|k\rangle \langle k|\}$ being the projective measurements on the subsystem $a$, which do not change the marginal state $\rho^{a}$ locally i.e., $\Pi ^{a}(\rho^{a})=\rho ^{a}$. If $\rho^{a}$ is non-degenerate, then the maximization is not required. 

\noindent \textit{Trace distance-based MIN}:
It is a well-known fact that the MIN based on the Hilbert-Schmidt norm is not a credible measure in capturing nonlocal attributes of a quantum state due to the local ancilla problem \cite{Chang2013}. A  natural way to circumvent this issue is defining MIN in terms of contractive distance measure. Another alternate form of MIN is based on trace distance \cite{HUTMIN},  namely, trace MIN (T-MIN)  which resolves the local ancilla problem  \cite{Piani2012}. 
It is defined as
\begin{equation}
N_1(\rho):= ~^{\text{max}}_{\Pi^a}\Vert\rho-\Pi^a(\rho)\Vert_1,
\end{equation} 
where $\Vert A \Vert_1 = \text{Tr}\sqrt{A^{\dagger}A}$ is the trace norm of operator $A$. Here also, the maximum is taken over all von Neumann projective measurements. For any $2\otimes2$ dimensional system, the closed formula of trace MIN $N_1(\rho)$ is given as 
\begin{equation}
N_1(\rho)=
\begin{cases}
\frac{\sqrt{\chi_+}~+~\sqrt{\chi_-}}{2 \Vert \textbf{x} \Vert_1} & 
 \text{if} \quad \textbf{x}\neq 0,\\
\text{max} \lbrace \vert c_1\vert,\vert c_2\vert,\vert c_3\vert\rbrace &  \text{if} \quad \textbf{x}=0,
\end{cases}
\end{equation}
where $\chi_\pm~=~ \alpha \pm 2 \sqrt{\tilde{\beta}} \Vert \textbf{x} \Vert_1 ,\alpha =\Vert \textbf{c} \Vert^2_1 ~\Vert \textbf{x} \Vert^2_1-\sum_i c^2_i x^2_i,\tilde{\beta}=\sum_{\langle ijk \rangle} x^2_ic^2_jc^2_k, \vert c_i \vert $ are the absolute values of $c_i$ and the summation runs over cyclic permutation of $\lbrace 1,2,3 \rbrace$ and $\text{x}_i$ are the components of vector $\textbf{x}$.

\section{The Model and Solution}

\label{model}
In order to understand the behaviour of quantum correlations in a physical system and the influence of intrinsic decoherence, we consider the Hamiltonian of a pair of spin-1/2 particles  with Heisenberg anisotropic interaction and DM interaction \cite{Dzyaloshinskii,Moriya} and is given as 
\begin{equation}
\mathcal{H} =\sum _{i = x,y,z} J_i \left(\sigma ^i_a \otimes \sigma ^i_b \right)+{\mathbf D}.\left( \bm{\sigma}_a\times  \bm{\sigma}_b\right) +(B+\lambda) \sigma _a^z+ (B-\lambda) \sigma_b^z,
\end{equation}
where $\sigma^i_k~(k = a,b ~ \text{and}~ i = x,y,z)$ are Pauli spin matrices, $J_i$'s are the anisotropic exchange coupling constant in respective directions, ${\mathbf D}$ is the DM vector which we choose to be along the $z$-axis, $B$ denotes the magnetic field and $\lambda$ denotes the degree of inhomogeneity in the magnetic field along the $z$-axis. In matrix notation we have\\
\begin{align}
\mathcal{H} = \begin{pmatrix}
 \frac{J_z}{2}+B & 0 & 0 & J_- \\
 0 & \lambda -\frac{J_z}{2} & J_+ + i D & 0 \\
 0 &  J_+ - i D & -\lambda -\frac{J_z}{2} & 0 \\
 J_- & 0 & 0 &  \frac{J_z}{2}-B
\end{pmatrix}
\label{ham}
\end{align}
with  $J_{\pm }=\frac{1}{2} \left(J_x\pm J_y\right)$. Eigenvalues and corresponding eigenvectors of the Hamiltonian are computed by solving Schr\"{o}dinger equation $\mathcal{H}|\phi\rangle = E|\phi\rangle$, which are
\begin{eqnarray}
E_{1,2}&=-\frac{J_z}{2}\pm \eta~~,~~|\phi_{1,2}\rangle &= N_\pm~~ \Bigl(\frac{ (\lambda\pm \eta )}{D+i J_+} |10\rangle  +  |01\rangle \Bigr) \nonumber\\
E_{3,4}&=\frac{J_z}{2}\pm \mu~~,~~~|\phi_{3,4}\rangle&= M_\pm~~\Bigl(\frac{B\pm \mu}{J_-} |00\rangle +|11\rangle \Bigr)
\label{eig}
\end{eqnarray}
with $\eta  = \sqrt{\lambda ^2+D^2+J_+{}^2} $ ; $\mu  = \sqrt{B^2+J_-{}^2}$ and normalization constants are 
\begin{equation}
N_{\pm } = \left(\frac{D^2+J_+{}^2}{2\eta\left(\eta ^2\pm \lambda  \right)}\right)^{1/2} ~~~~~\text{and}~ ~~~~~ \nonumber \\
M_{\pm } =  \left(\frac{J_-{}^2}{2\mu\left(\mu ^2 \pm B \right)}\right)^{1/2}.
\end{equation}
For $J_x=J_y=J_z$ and $D_z=B=\lambda=0$ the eigenvectors are reduced to maximally entangled Bell states.

Milburn \cite{Milburn} has proposed a simple modification of standard quantum mechanics based on an assumption that on sufficiently short time steps the system does not evolve continuously under unitary evolution but rather in a stochastic sequence of identical unitary transformations. This assumption leads to a modification of the Schrödinger equation which contains a term responsible for the decay of quantum coherence in the energy eigenstate basis, without the intervention of a reservoir and therefore without the usual energy dissipation associated with normal decay. Milburn obtained the following master equation \cite{Milburn}:
\begin{equation}
\frac{d\rho(t)}{dt} = - i[\mathcal{H},\rho(t)]-\frac{\gamma}{2}[\mathcal{H},[\mathcal{H},\rho(t)]],
\label{milburn1}
\end{equation}
where $\mathcal{H}$ is the Hamiltonian of the system, $\rho(t)$ indicates the state of the system and $\gamma$ is the intrinsic decoherence parameter (mean frequency of the unitary step). First term on the right--hand side of Eq. (\ref{milburn1}) generates a coherent unitary time evolution of the system, while the second term, which does not commute with the Hamiltonian represents the decoherence effect on the system. The formal solution of the above equation can be written in operator-sum representation using Kraus operators $M_l$ as
\begin{equation}
\rho(t)= \sum^{\infty}_{l=0}  M_l(t) \rho(0)M_l{}^{\dagger}(t),
\end{equation}
where $\rho(0)$ is the initial density operator of the system and $M_l(t)$ is defined as
\begin{align*}
M_l(t) = \frac{(\gamma t)^{\frac{l}{2}}}{\sqrt{ l!}} \mathcal{H}^l \text{exp}(-i\mathcal{H}t) \text{exp}(-\frac{\gamma t}{2}\mathcal{H}^{2}).
\end{align*}
With this,  the evolved state can be written as 
\begin{equation}
\rho(t) = \sum_{m,n} \text{exp}\Bigl[-\frac{\gamma t}{2}(E_{m}-E_{n})^{2}- i(E_{m}-E_{n})t\Bigr] \langle\phi_{m}\vert\rho(0)\vert\phi_{n}\rangle \vert\phi_{m}\rangle\langle\phi_{n }\vert,
\label{sol}
\end{equation}
where $E_{m,n}$ and $|\phi_{m,n}\rangle $ are the eigenvalues and the corresponding eigenvectors of Hamiltonian $\mathcal{H}$ the system respectively.
\section{Dynamics of Quantum Correlations}
\label{Dynamics}
In what follows, we study the influence of intrinsic decoherence  on quantum correlations captured by the entanglement (quantified by concurrence) and measurement-induced nonlocality based on Hilbert-Schmidt norm and trace distance. In order to study the dynamical behaviour of quantum correlation, we will consider the X--state as an initial which encompasses a large number of mixed states \cite{Yurischev}:
\begin{align}
\rho(0) = \begin{pmatrix}
	a&0&0&\omega \\
	0&b&z&0 \\
	0&z&c&0 \\
	\omega&0&0&d 
\end{pmatrix},
\end{align}
where the diagonal entries are real, non-negative and satisfy the normalization condition $\text{Tr}(\rho(0))=a+b+c+d=1$. 

The evolution of X-state under the Hamiltonian (\ref{ham}) also retains the form of X--state. Here, we find  $\rho(t)$ from the solution of Master equation as given in Eq. (\ref{sol}). Using the Eqs. (\ref{eig}) and (\ref{sol}) we compute the time evolved density matrix as
\begin{align}
\rho(t) = \begin{pmatrix}
	\rho_{11}(t)&0&0&\rho_{14}(t) \\
	0&\rho_{22}(t)&\rho_{23}(t)&0 \\
	0&\rho_{23}^*(t)&\rho_{33}(t)&0 \\
	\rho_{14}^*(t)&0&0&\rho_{44}(t) 
\end{pmatrix}
\end{align}
with the diagonal entries 
\begin{align*}
\rho_{11} &= \frac{1}{2\mu ^2}\Bigl((a(B^2 + \mu ^2)+ 2 B w J_- + d J_-^2)  + e^{-2 \gamma t \mu^2} \text{cos}[2 \mu t]((a-d)J_-^2 -2 B w J_-)\Bigr), \nonumber \\
\rho_{22} &= \frac{1}{2 \eta ^2}\Bigl((c(D^2+J_+^2)+b(\eta ^2+\lambda ^2)+2 z \lambda  J_+)+ e^{-2 \gamma t \eta ^2}(2 D z \eta  \text{sin}[2\eta t]\nonumber\\
          & + \text{cos}[2 \eta t]((b-c)(D^2+J_+^2)-2 z \lambda  J_+))\Bigr), \nonumber\\
\rho_{33} &= \frac{1}{2 \eta ^2}\Bigl((c(\eta ^2+\lambda ^2)+b(\eta ^2-\lambda ^2)-2 z \lambda  J_+)-e^{-2 \gamma t \eta ^2}(2 D z \eta \text{sin}[2 \eta t]+ cos[2 \eta t] \nonumber \\
          & ((b-c)(D^2+J_+^2)-2 z \lambda  J_+))\Bigr), \nonumber\\
\rho_{44} &=\frac{1}{2\mu ^2}\Bigl((aJ_-^2 -2 B w J_-+d(B^2+\mu ^2))- e^{-2 \gamma t \mu ^2} \text{cos}[2\mu t]((a-d)J_-^2-2 B w J_-)\Bigr),\nonumber\\
\end{align*}
and off-diagonal elements are 
\begin{align}
\rho_{14}=\rho^*_{41} &=\frac{1}{2\mu ^2}\Bigl(((a-d)B J_- + 2w J_-^2)-e^{-2\gamma t \mu ^2}((a-d)J_- -2 B w)\nonumber\\                                    &(B~ \text{cos}[2 \mu t]- i \mu~ \text{sin}[2 \mu t])\Bigr),\nonumber\\
\rho _{23}=\rho^*_{32}&=\frac{1}{2 \eta ^2}\Bigl( i(D - iJ_+)((b-c )\lambda +2z J_+)
+\frac{e^{-2 \gamma t \eta ^2}}{\eta ^2-\lambda ^2}(\eta~ \text{sin}[2\eta t](D-i J_+)(-(b-c)\nonumber\\
          &(D^2+J_+^2)-2 i D z \lambda + 2 z \lambda  J_+)+ \text{cos}[2 \eta t](i D+J_+)(-(b-c)\lambda(D^2+J_+^2)\nonumber\\& -2 i D z \eta ^2+2 z \lambda ^2 J_+))\Bigr).\nonumber    
\end{align}

The concurrence of the time evolved state is 
\begin{equation}
C(\rho(t))=2 \text{max} \lbrace 0,\vert \rho _{14} \vert -\sqrt{(\rho_{33}\rho_{22})},\vert \rho _{23} \vert -\sqrt{(\rho_{44}\rho_{11})}\rbrace,
\end{equation}
and MINs are computed as 
\begin{align}
N_2(\rho(t)) &=  2 (\vert \rho_{23} \vert{}^2 + \vert \rho_{14} \vert{}^2)~~~~~~~ \text{and}~~~~~~
N_1(\rho(t)) &=  2 (\vert \rho_{23} \vert + \vert \rho_{14} \vert).
\end{align}

For the detailed investigations, we choose a special kind of X-state as  
\begin{equation}
\rho(0) = p|\Phi\rangle \langle \Phi| + (1-p) \frac{\mathds{1}}{4},
\label{state}
\end{equation}
where $p$ is the probability of being in state $|\Phi\rangle \langle \Phi|$, with $0 \leq p \leq 1$. If $p=0$, the state is a maximal mixture of orthonormal bases i.e., $\rho(0) = \frac{I}{4}$ and if  $p=1$ the state becomes a pure state. For our purpose, we consider the above state with the pure state $|\Phi\rangle=\frac{1}{\sqrt{2}} (|01\rangle  +  |10\rangle ) ~~\text{or}~~ |\Phi\rangle=\frac{1}{\sqrt{2}}(|00\rangle  +  |11\rangle )$.  We shall note that the evolution of this  state under the Hamiltonian (\ref{ham}) retains the form of initial state.

\begin{figure*}[!ht]
\centering\includegraphics[width=0.6\linewidth]{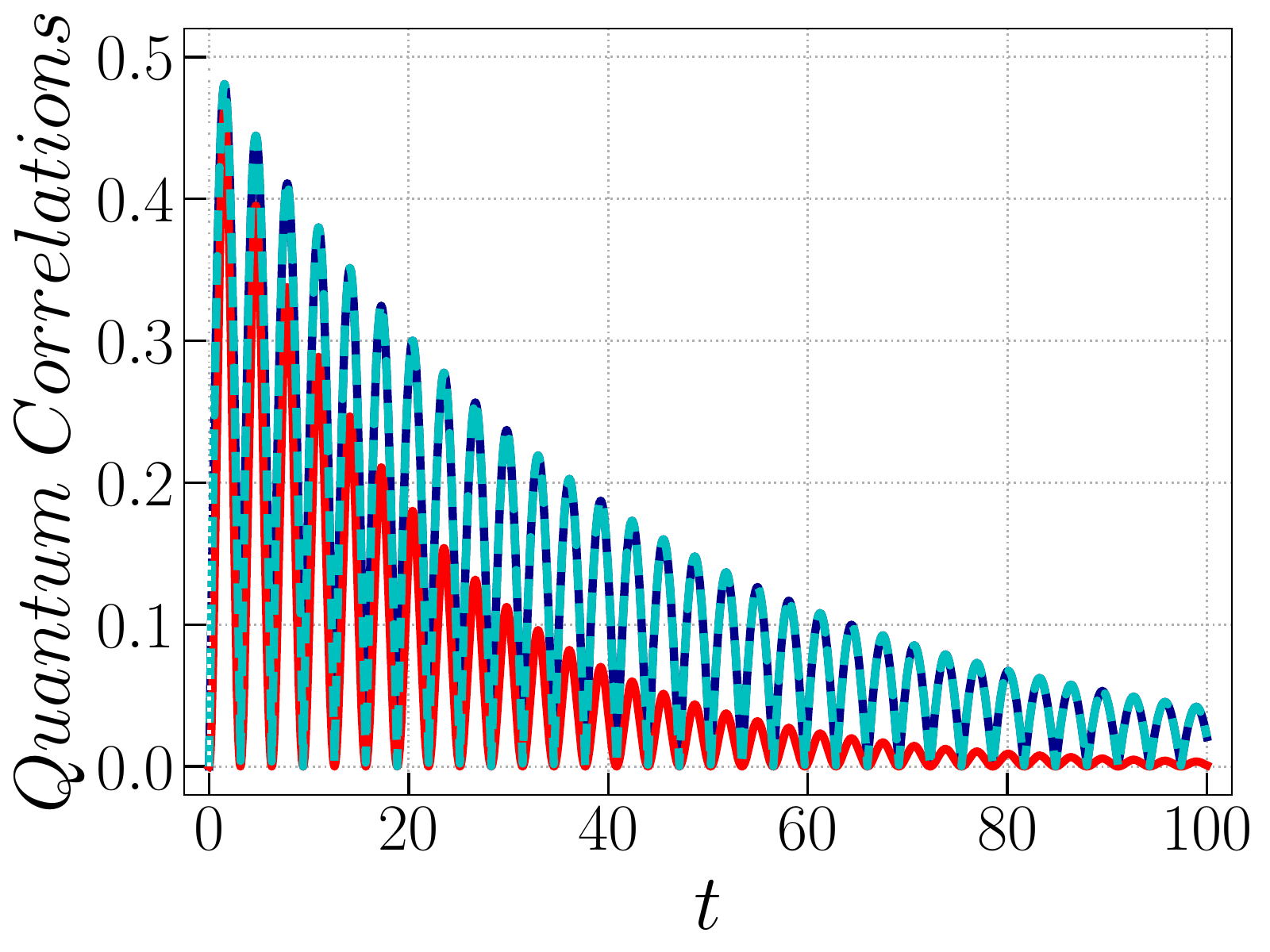}
\centering\includegraphics[width=0.6\linewidth]{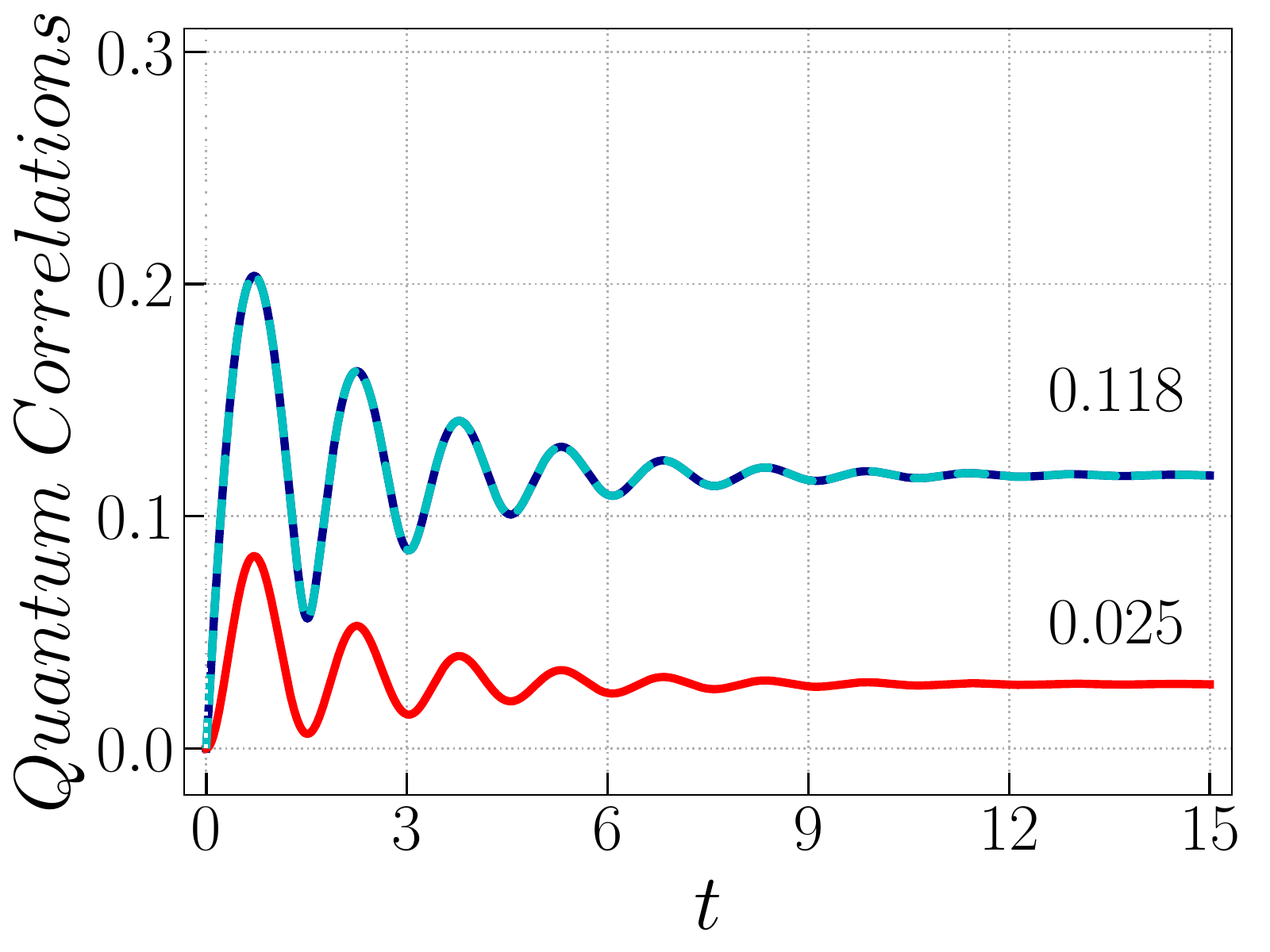}
\caption{(color online) Dynamical behaviors of concurrence (blue), Hilbert-Schmidt (red) and trace distance (dashed) based MIN of the state $\rho(0)=|00\rangle \langle 00|$ as a function of time for the parameters $J_+=1, J_=0.5, \gamma=0.05, p=1$ with (top) $\lambda=D=B=0$ (bottom) $\lambda=0.5, D=3, B=1$.}
\label{p1}
\end{figure*}

To explore the effects of intrinsic decoherence, we study the dynamics of pure state from the perspective of quantum correlation. For pure state dynamics, we identify the simple relation between the entanglement and MINs. They are 
\begin{equation}
C(\rho)=\frac{1}{2} N_1(\rho) ~~~~~~~ \text{and}~~~~~ N_2(\rho)=\frac{1}{2}C(\rho)^2. \label{Relation}
\end{equation}

For our investigations, first we choose the initial state  with the following initial conditions: $a=(1+p)/4$, $b=c=d=(1-p)/4$, $\omega=0$ and $z=0$. Here, we set $p=1$, and the initial state is a product state $\rho(0)=|00\rangle \langle 00|$, i.e., at $t=0$ the correlation between the qubits are zero. As time increases, the correlation between the qubits generated by the time evolution of the state under the Hamiltonian (\ref{ham}), implying that the spin–spin interaction is responsible for the induced correlation between qubits, which are shown in Fig. (\ref{p1}). Both the entanglement and MIN measures are oscillating with period $\pi/\mu$ and the damping factor $\exp[-2 \gamma t \mu ^2]$  cause decay of amplitude of the oscillation exponentially, reaches a steady state after a long time. The role of DM interaction and magnetic fields on the dynamics of quantum correlations are shown in Fig. (\ref{p1}). It is a well-known fact that the DM interaction enhances the strength of entanglement and MIN. Further, the applications of the magnetic field decrease the quantum correlations and the state evolves into finite steady state values at the earlier time compare to the zero-field case. The steady state value for this case  are $C(\rho(\infty ))=0.236$ and $N_2(\rho(\infty ))=0.025$. We consider another initial product state of interacting qubits such as $|\Phi\rangle= |0\rangle\otimes|1\rangle$.  As time increases, the spin-spin interaction described by the Hamiltonian induces the correlation between the qubits by the time evolution of the state. Here also we can observe the similar effects of the previous case of product state. From the above observations, it is understood that the intrinsic decoherence can  decrease the correlation between the correlated spins asymptotically.  

\begin{figure*}[!ht]
\centering\includegraphics[width=0.6\linewidth]{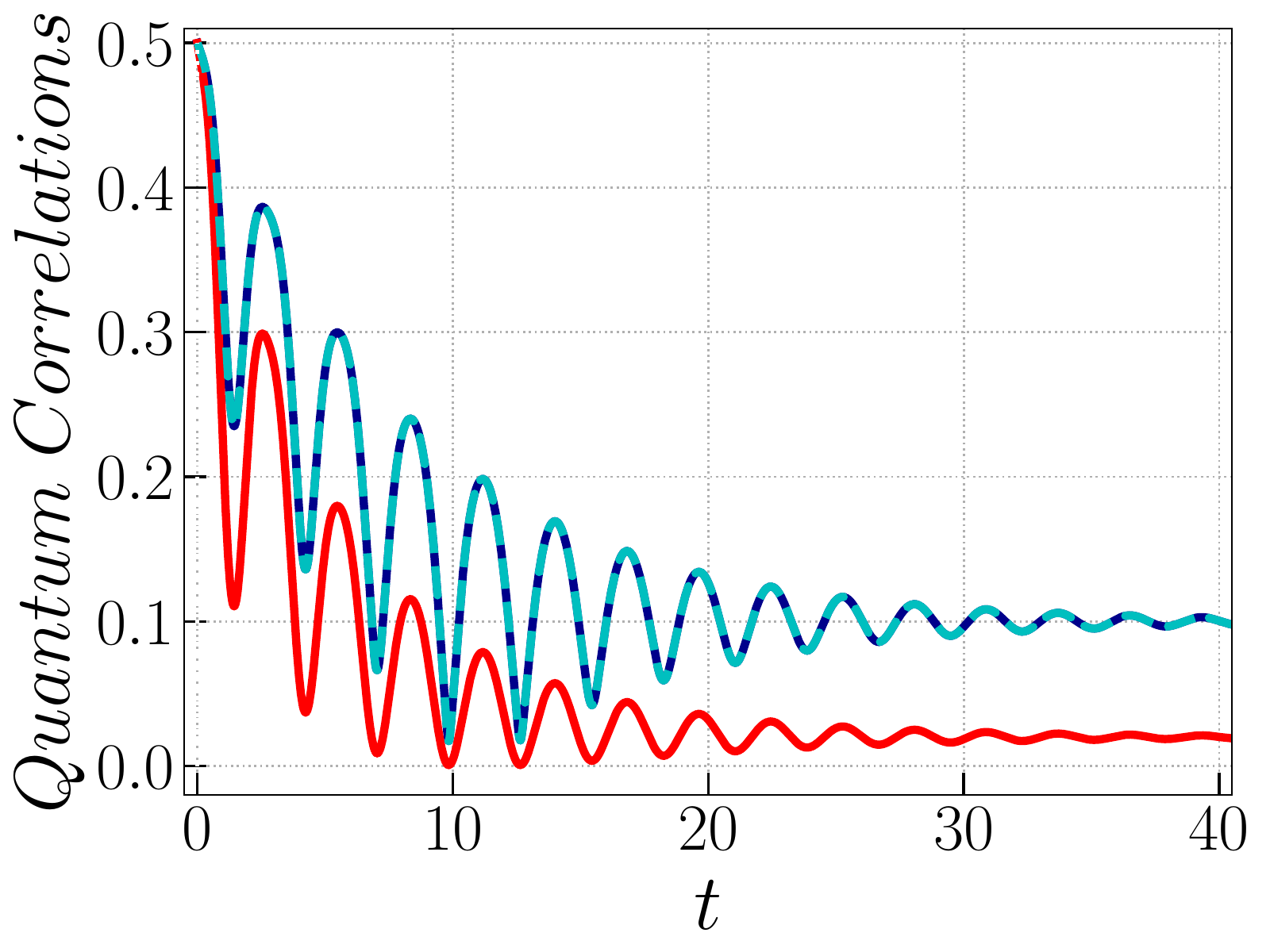}
\centering\includegraphics[width=0.6\linewidth]{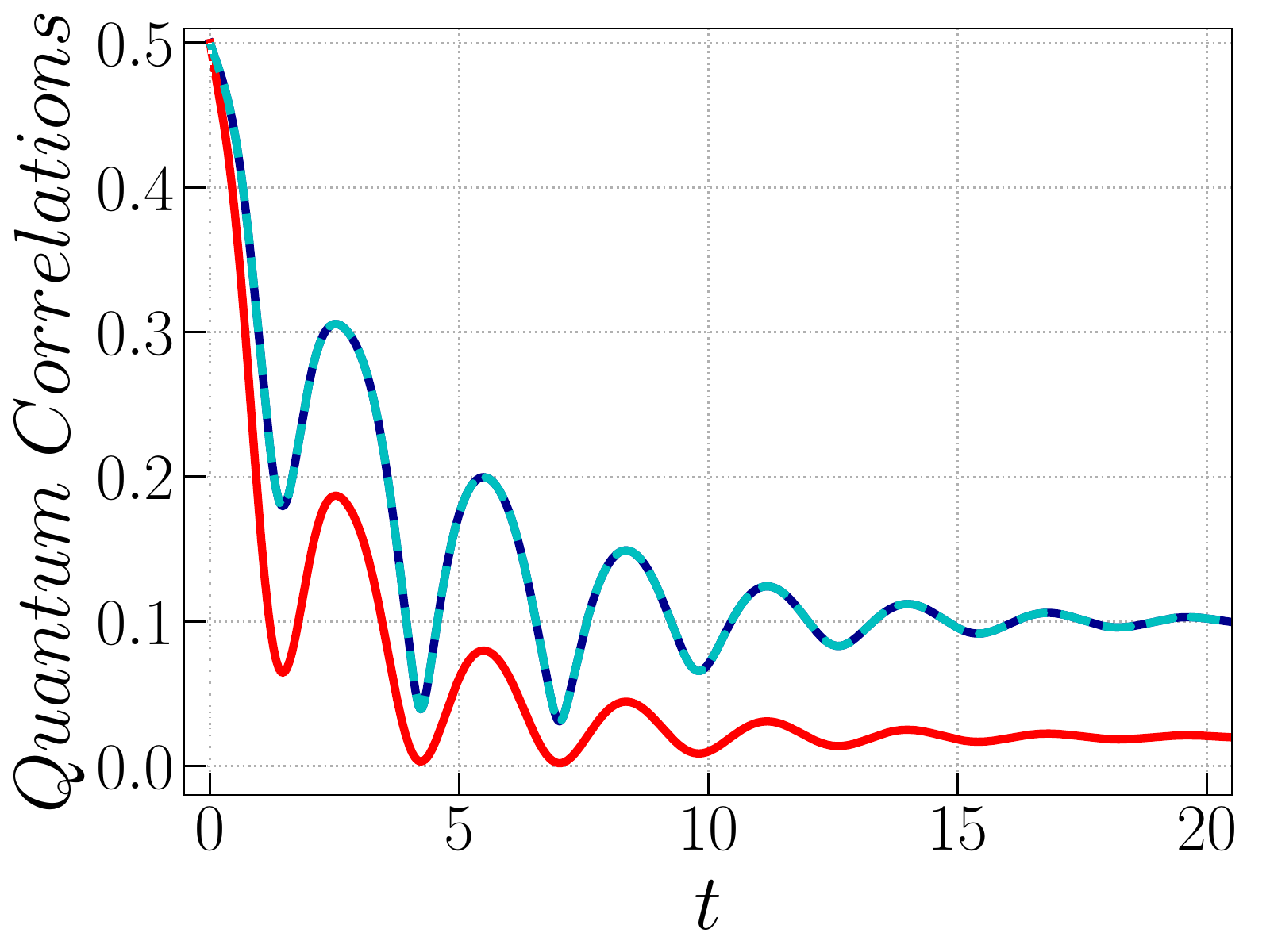}
\centering\includegraphics[width=0.6\linewidth]{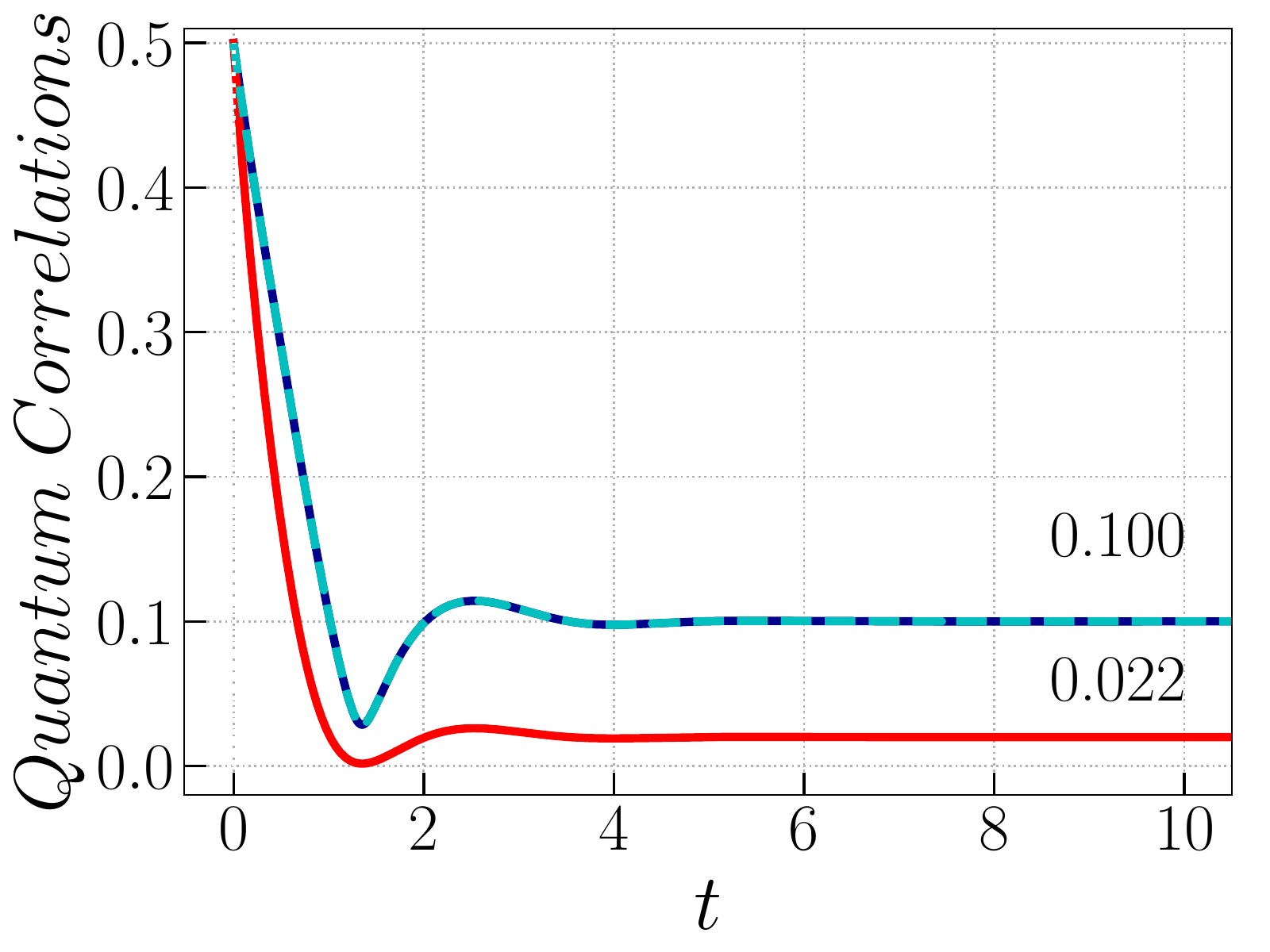}
\caption{(color online) Dynamical behaviors of concurrence (blue), Hilbert-Schmidt (red) and trace distance (dashed) based MIN of the state  $|\Phi\rangle=\frac{1}{\sqrt{2}}(|00\rangle+|11\rangle ) $ as a function of time for the parameters $J_+ = 1, J_-= 0.5, p = 1, \lambda=0.5, D=1, B=1$ with (top) $\gamma=0.05$ (middle)
$\gamma=0.1$ and  (bottom) $\gamma=0.3$.}
\label{fig2}
\end{figure*}

Next, we study the dynamics of pure entangled state $\rho(0)=|\Phi\rangle \langle \Phi|$ with $|\Phi\rangle =\frac{1}{\sqrt{2}}(|00\rangle+|11\rangle )$. Setting $p=1$ and the initial state is the maximally entangled state. Here also we obtain the relation given in Eq. (\ref{Relation}). At time $t = 0$, the concurrence and MIN are maximum. As time increases, the correlation decrease due to the damping factor $\exp[-2 \gamma t \mu ^2]$  with a period of oscillation of $\frac{\pi}{\mu}$, which are illustrated in Fig. (\ref{fig2}). For  $\gamma=0.05$, the quantum correlation measures are wiggles and in the asymptotic limit $t\rightarrow\infty$, 
the concurrence and MIN reaches the steady state value. Further, Fig. (\ref{fig2}) shows the role of higher $\gamma=0.1 ~\&~ 0.5$ values. The increment of the decoherence parameter $\gamma$ causes rapid decay in correlation measures and reaches a steady state as earlier. In the asymptotic limit i.e., $t\rightarrow\infty$, irrespective of $\gamma$ values, both concurrence and MIN reaches a constant value. In other words, for $\gamma=0.5$, the concurrence and MIN of the steady state are constant and the values are $\mathcal{C}(\rho(\infty))= 0.2$ and $\mathcal{N}_2(\rho(\infty)) = 0.025$. Any change in the value of $\gamma$ will affect only the oscillation part of dynamics, and the asymptotic limit doesn't depend upon this $\gamma$ but it depends on the initial state of the system. Here also we may note that the DM interaction and coupling constant strengthen the correlation and the intervention of the magnetic field decreases the correlation between the qubits. 

\begin{figure*}[!ht]
\centering\includegraphics[width=0.6\linewidth]{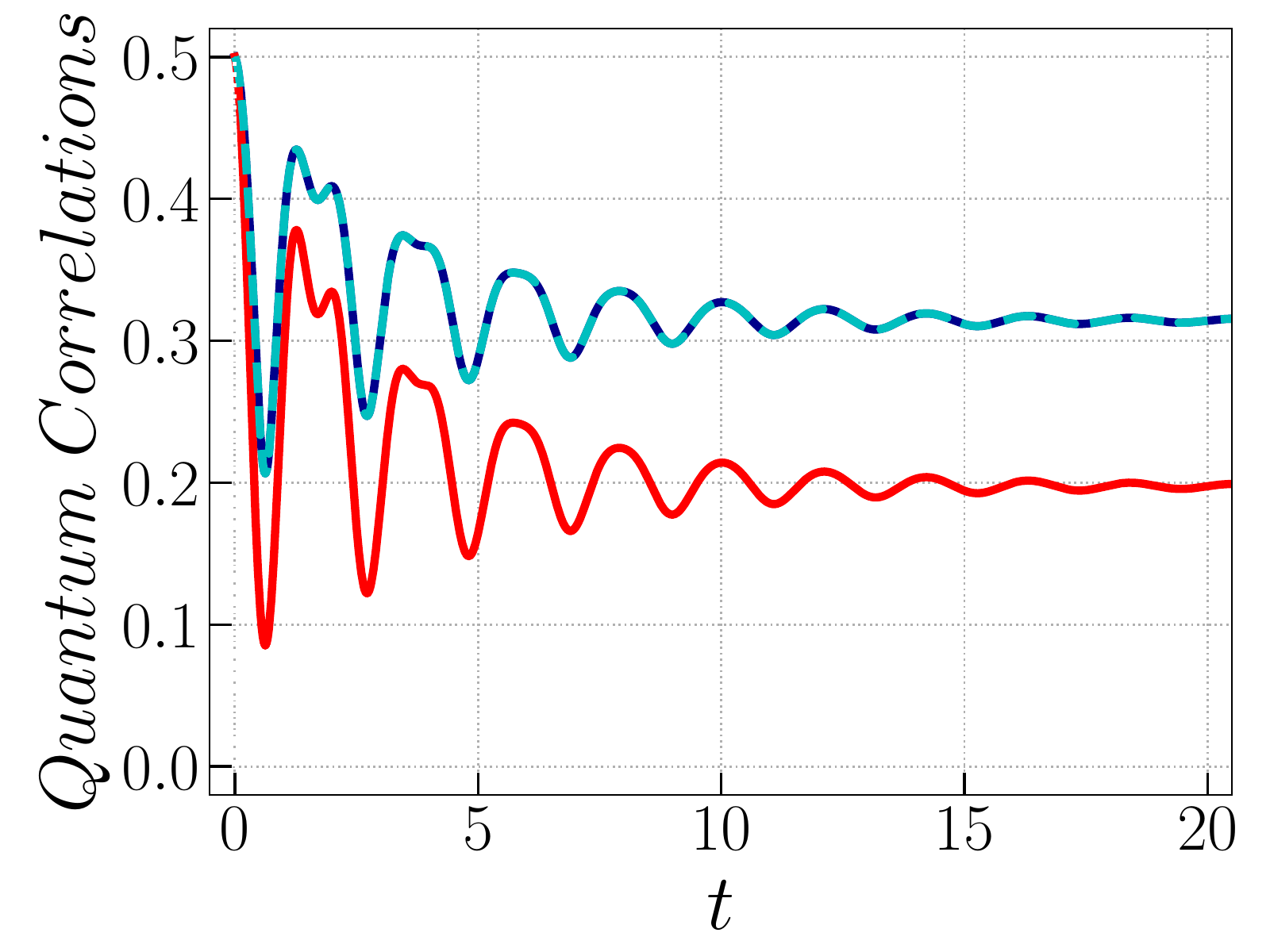}
\centering\includegraphics[width=0.6\linewidth]{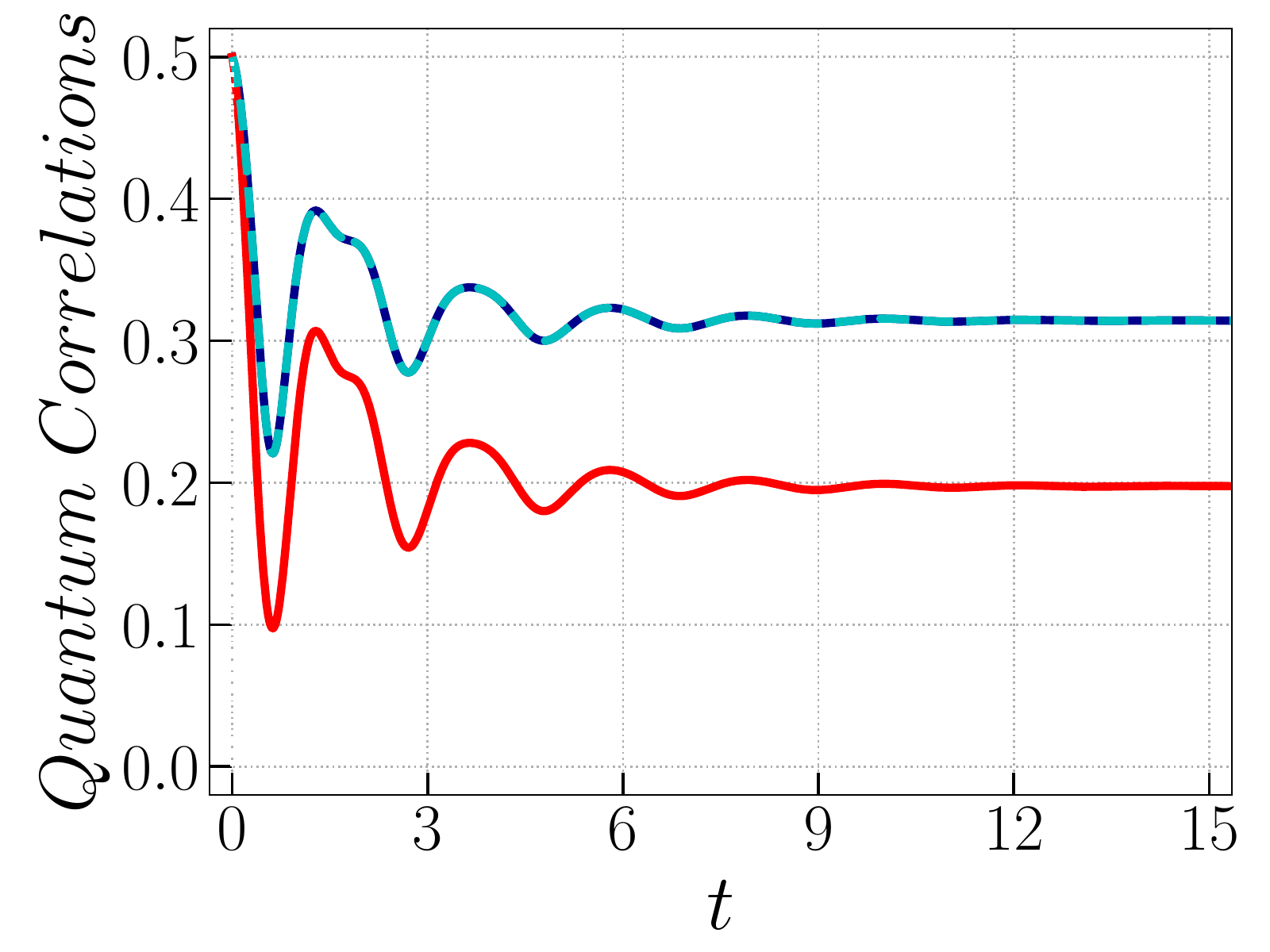}
\centering\includegraphics[width=0.6\linewidth]{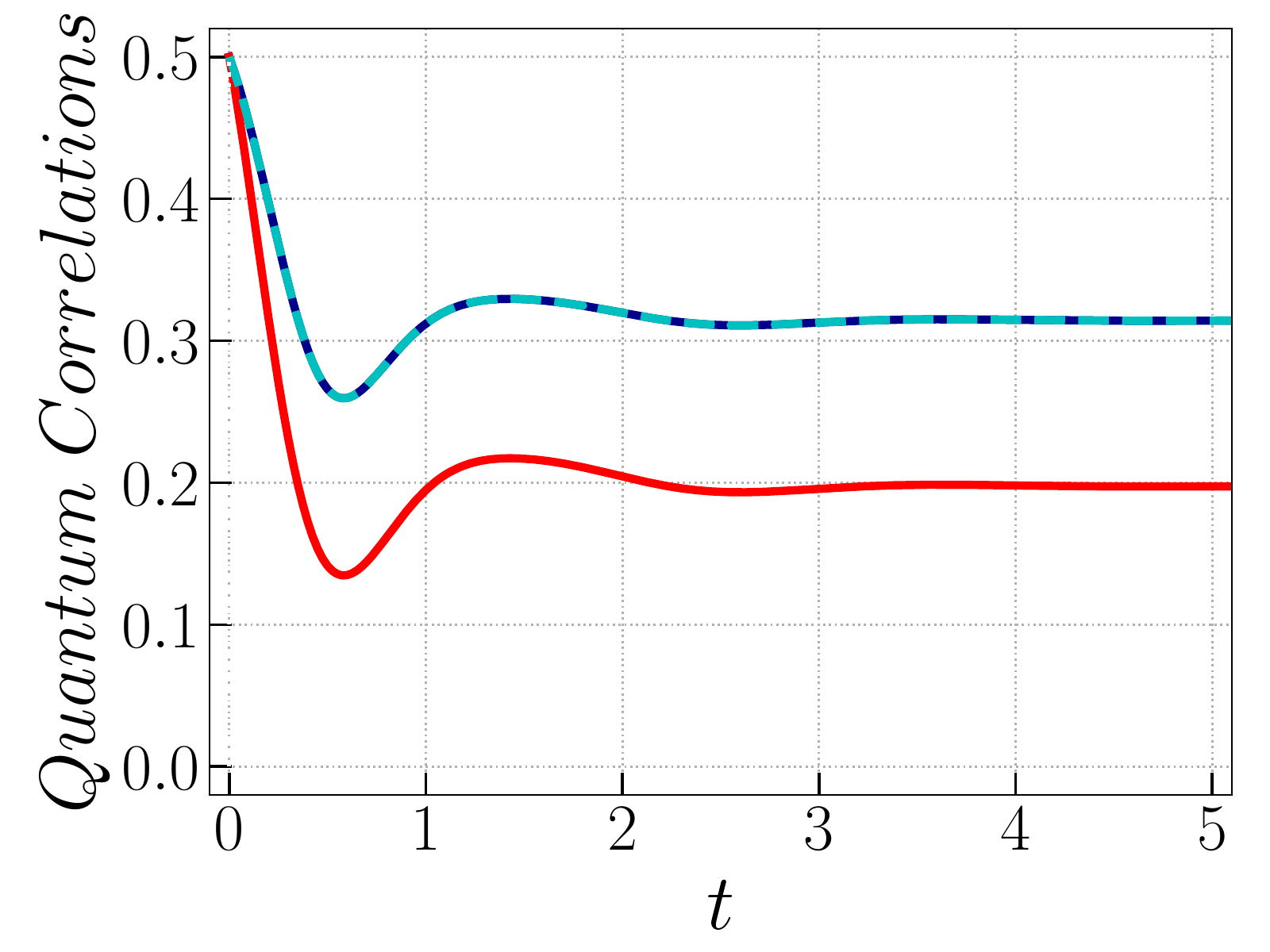}
\caption{(color online) Dynamical behaviors of concurrence (blue), Hilbert-Schmidt (red) and trace distance (dashed) based MIN of the state  $|\Phi\rangle=\frac{1}{\sqrt{2}}(|01\rangle+|10\rangle ) $ as a function of time for the parameters $J_+ = 1, J_-= 0.5, p = 1, \lambda=0.5, D=1, B=1$ with (top) $\gamma=0.05$ (middle)
$\gamma=0.1$ and  (bottom) $\gamma=0.3$.}
\label{fig3}
\end{figure*}

To enhance our understanding, we now consider another pure maximally entangled state $\rho(0)=|\Phi\rangle \langle \Phi|$ with $|\Phi\rangle =\frac{1}{\sqrt{2}}(|01\rangle+|10\rangle )$. The initial conditions are $a=d=(1-p)/4$, $b=c=(1+p)/4$, $\omega=0$ and $z=p/2$  with $p=1$. The relation between the quantum correlation measures is given in Eq. (\ref{Relation}). Fig. (\ref{fig3}) display the effects of intrinsic decoherence on dynamics concurrence and MIN quantities. At time $t=0$, measures  show the maximum correlation. Further, at a later time, due to the damping factor $\exp[-2 \gamma t \eta^2]$, all the quantum correlation measures are decreasing with the increase of time. For $\gamma=0.05$, the quantum correlation measures are wiggles and in the asymptotic limit, all the measures reach a steady state value. If we consider higher values of $\gamma$, one can observe the similar effects observed in the previous case.  Here also the steady state value independent of intrinsic decoherence and depends on only the initial conditions.

\begin{figure*}[!ht]
\centering\includegraphics[width=0.6\linewidth]{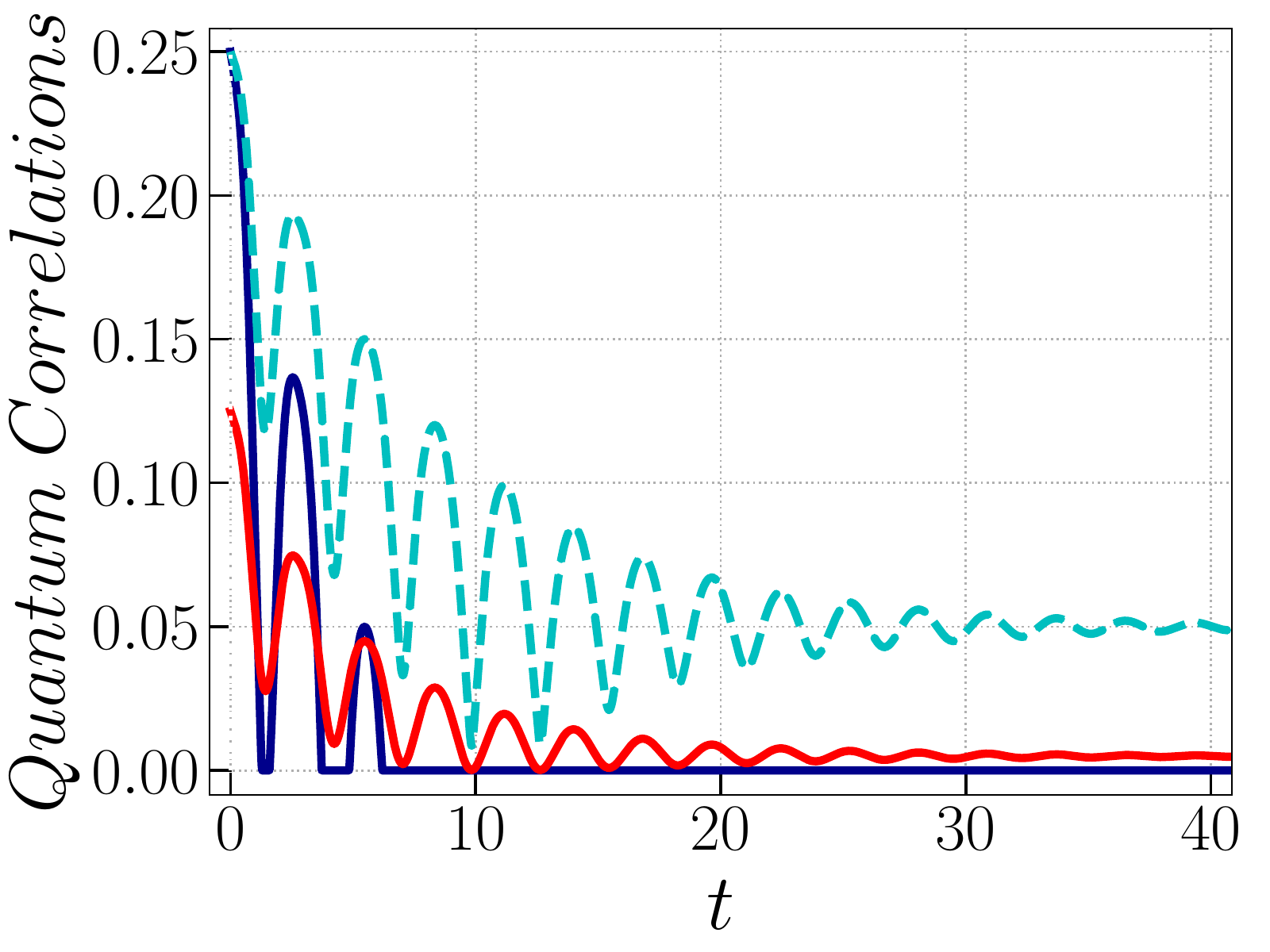}
\centering\includegraphics[width=0.6\linewidth]{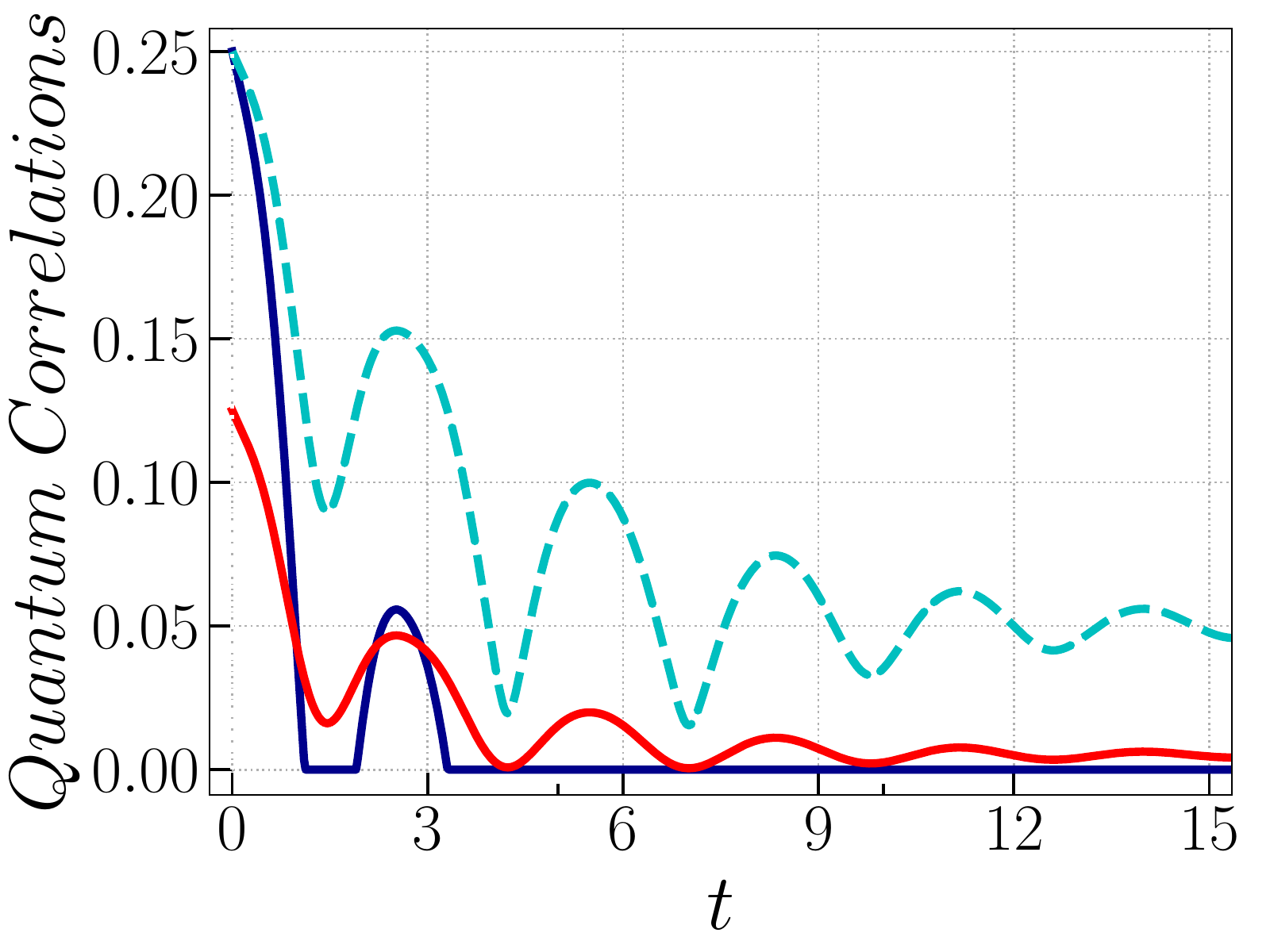}
\centering\includegraphics[width=0.6\linewidth]{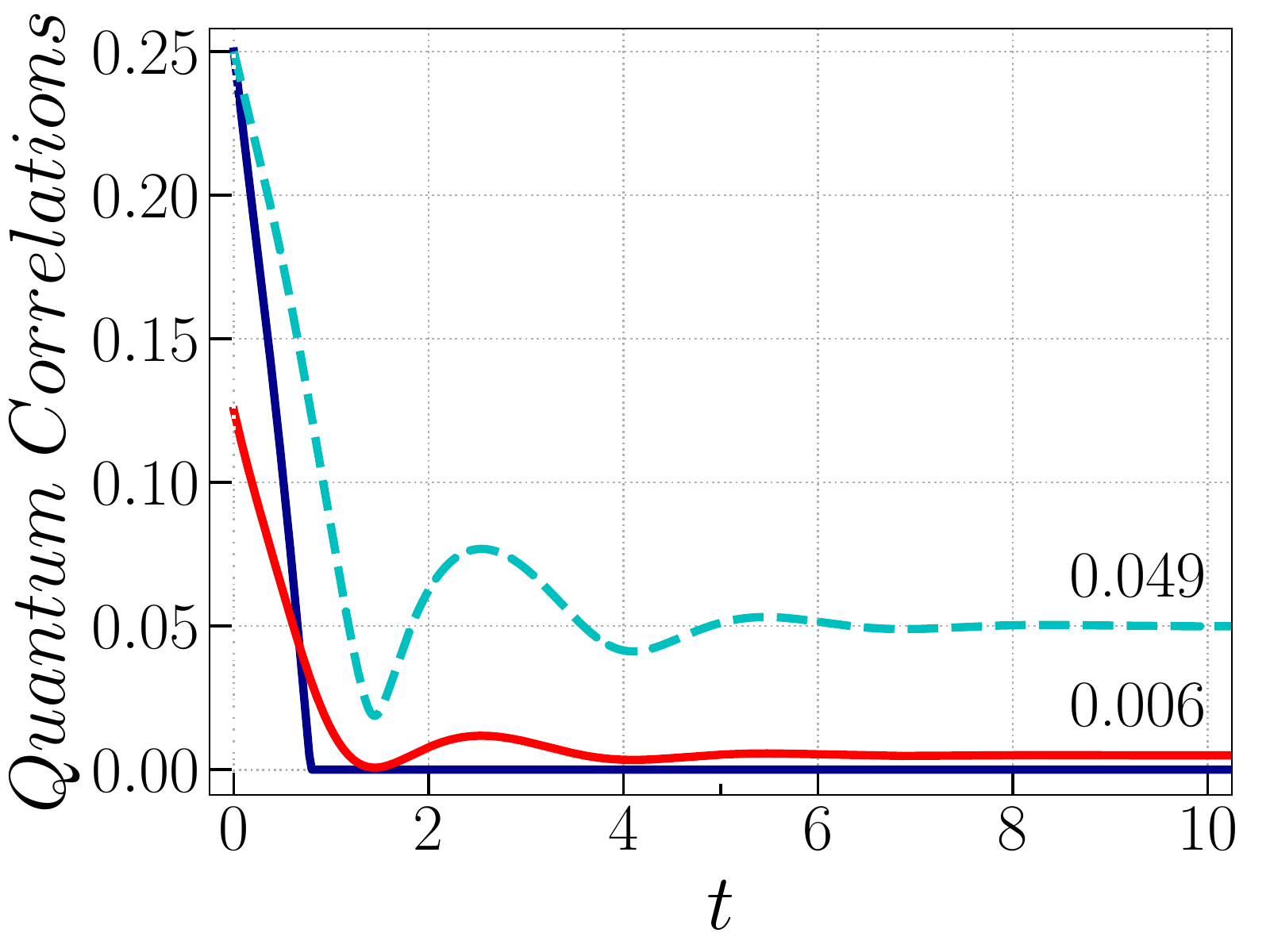}
\caption{(color online) Dynamical behaviors of concurrence (blue), Hilbert-Schmidt (red) and trace distance (dashed) based MIN of the state $\rho(0)$ in Eq. (\ref{state}) with  $|\Phi\rangle=\frac{1}{\sqrt{2}}(|00\rangle+|11\rangle ) $ as a function of time for the parameters $J_+ = 1, J_-= 0.5, p = 0.6, \lambda=0.5, D=1, B=1$ with (top) $\gamma=0.05$ (middle)
$\gamma=0.1$ and  (bottom) $\gamma=0.3$. }
\label{fig4}
\end{figure*}

To examine the dynamics of quantum correlation of mixed state, we consider the initial non--separable state i.e, a mixture of entangled state with $|\Phi\rangle = \frac{1}{\surd 2} (|00\rangle +  |11\rangle)$ for our dynamics with the following initial conditions: $a=d=(1+p)/4, b=c=(1-p)/4,~ z=p/2, \omega=0$ and $p=0.6$. In this analysis also we observe that the dynamics of quantum correlation is periodic in time with periodicity $\frac{\pi}{\mu}$ and decay with time. The concurrence is maximum when $t=0$, as time increases, the concurrence decreases and undergoes death and revival which are shown in Fig. (\ref{fig4}). Interestingly, after some time entanglement between the qubits remains zero and is known as the sudden death of entanglement \cite{sudden}. On the other hand, both the Hilbert-Schmidt norm and trace distance-based MIN also decreases with time from the maximal value. In the asymptotic limit, MINs are non-zero even in the entanglement. It implies that the separable state also possesses the quantum correlation. This observation reveals that the presence of quantum correlation beyond entanglement in the absence of concurrence. Further, the role of intrinsic decoherence is shown in the remaining figures and one observes the similar effects which are observed in the previous example. For higher values of $\gamma$  entanglement decrease rapidly and also exhibits sudden death. The region of non-zero entanglement decreases with the increase of intrinsic decoherence.  On the other hand, MINs decrease slowly compared to entanglement and reaches a steady state after a sufficient time and the steady state value is $0.049$. In the case of   $|\Phi\rangle = \frac{1}{\surd 2} (|01\rangle +  |10\rangle)$. on the dynamics of concurrence and MIN, one can observe similar effects.

\begin{figure*}[!ht]
\centering\includegraphics[width=0.6\linewidth]{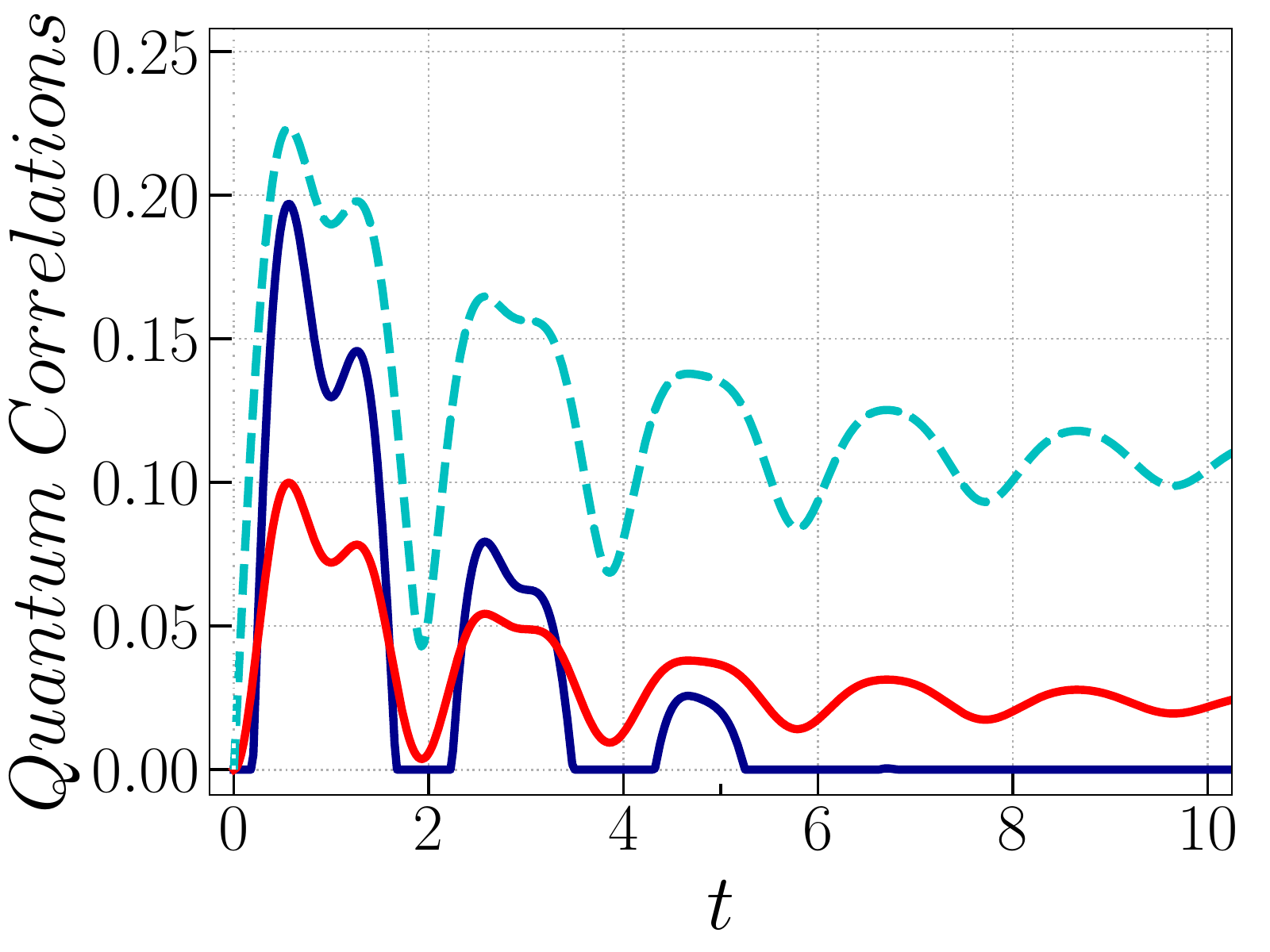}
\centering\includegraphics[width=0.6\linewidth]{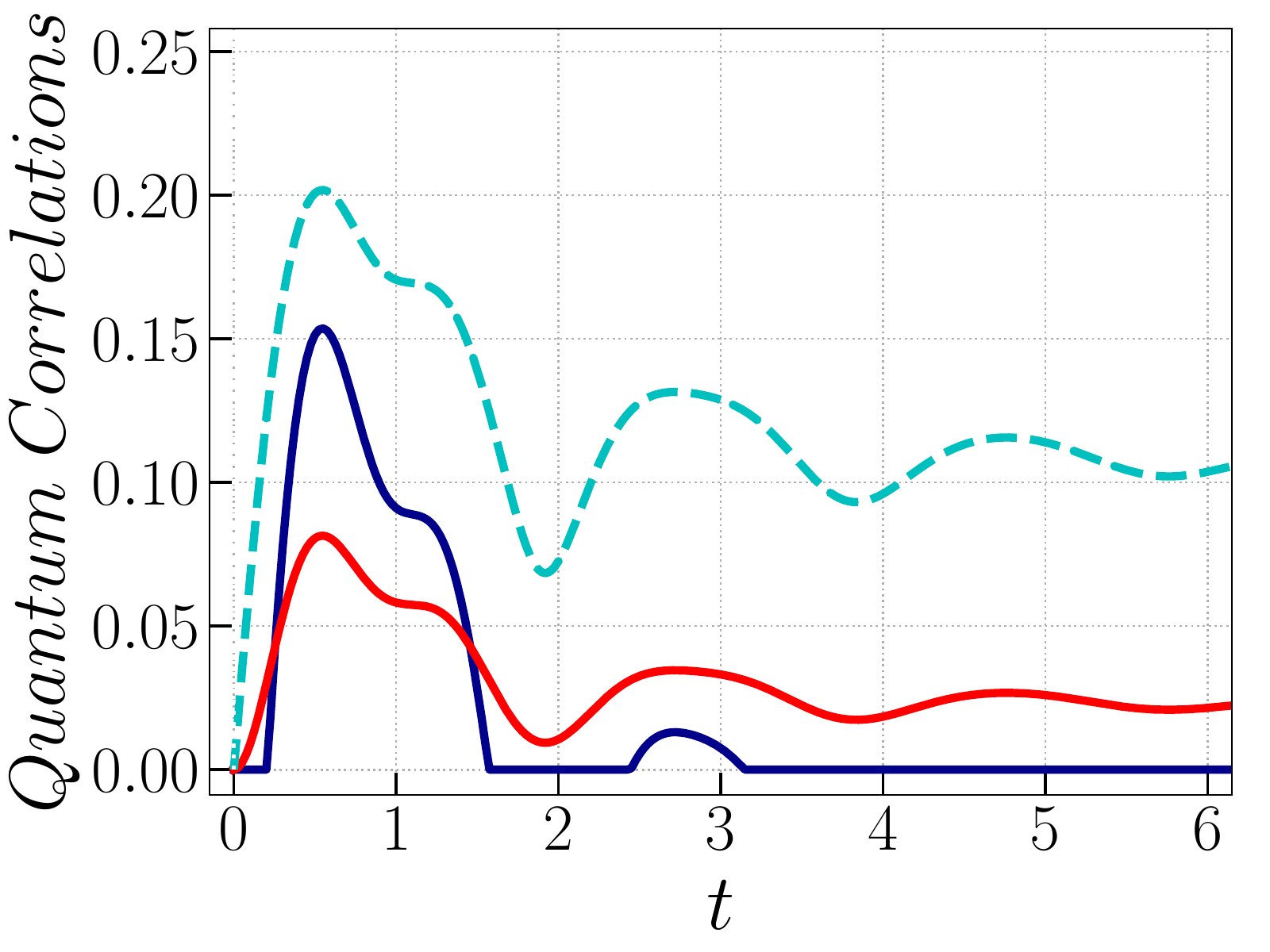}
\centering\includegraphics[width=0.6\linewidth]{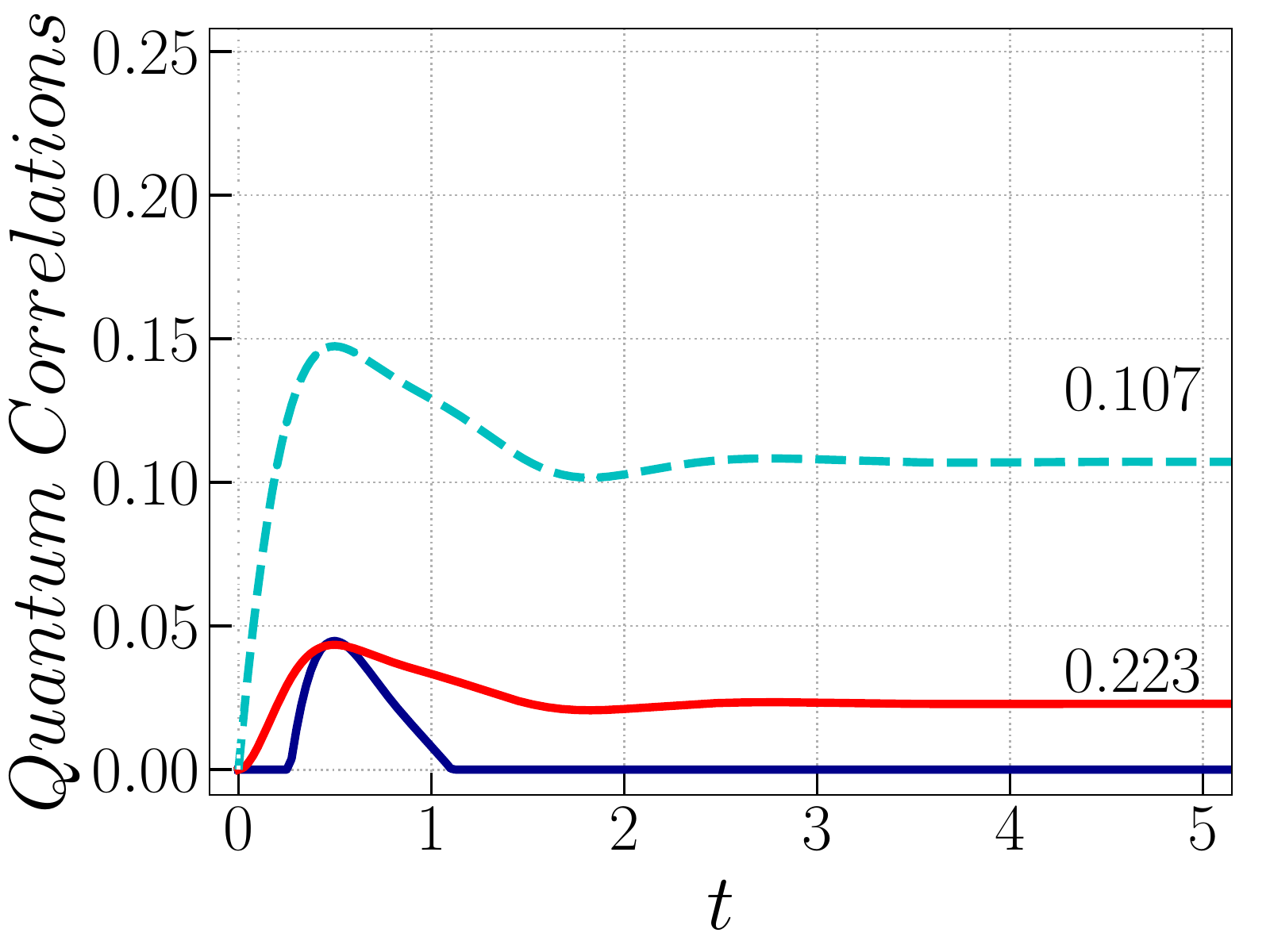}
\caption{(color online)  Dynamical behaviors of concurrence (blue), Hilbert-Schmidt (red) and trace distance (dashed) based MIN of the state $\rho(0)$ in Eq. (\ref{state}) with $|\Phi\rangle=|00\rangle$ as a function of time for the parameters $J_+ = 1, J_-= 0.5, p = 0.6, \lambda=0.5, D=1, B=1$ with (top) $\gamma=0.05$ (middle) $\gamma=0.1$ and  (bottom) $\gamma=0.3$.}
\label{fig5}
\end{figure*}




For further understanding, we consider the separable state with $|\Phi\rangle = |10\rangle~\text{or}~|00\rangle$. At time $t=0$, all the quantum correlations are zero. Here also the spin-spin interaction is responsible for the generation of nonlocal attributes which are illustrated in Fig. (\ref{fig5}). Similar to the entangled state, the dynamics of the separable state also exhibits sudden death for all values of $\gamma$. The other companion quantities are non-zero in the asymptotic limit and more robust to intrinsic decoherence than the entanglement.  This also implies to a fact that the absence of entanglement does not necessarily indicate the absence of nonlocality. It is worth mentioning that MIN is also robust against external decoherence \cite{Muthu2}.

\section{Conclusions}
\label{Concl}
In conclusion, we have examined the influence of intrinsic decoherence on various correlations in the model of two spin-1/2 qubits. While considering the pure state as an initial state, we obtain a simple relation between the concurrence and MINs. The results show that despite the intrinsic decoherence all the correlations reach their steady state values after exhibiting some oscillations. The Dzyaloshinskii–Moriya (DM) interaction strengthens the quantum correlations and the applications of the magnetic field decrease the quantum correlations. For the initial mixed state, we observe that the dynamics under intrinsic decoherence cause sudden death in entanglement, while MIN quantities are more robust than entanglement. This indicates the presence of nonlocality (in terms of MIN) even in the absence of entanglement between the subsystems.  This also implies to a fact that the absence of entanglement does not necessarily indicate the absence of nonlocality.

Further, our investigations emphasize that efficient  information processing based on the measurement-induced nonlocality offer more resistance to the effect of intrinsic decoherence and are completely different from that of entanglement.

%
%

\begin{acknowledgements}

This work has been financially supported by the CSIR EMR Grant No. 03(1444)/18/EMR-II.
\end{acknowledgements}



\end{document}